\documentclass[conference]{IEEEtran}
\IEEEoverridecommandlockouts
\usepackage{cite}
\usepackage{amsmath,amssymb,amsfonts}
\usepackage{graphicx}
\usepackage{textcomp}
\usepackage{tabularx}
\usepackage{multirow}
\usepackage{lipsum}  
\usepackage[table,xcdraw]{xcolor}
\usepackage{hyperref} 
\usepackage{cleveref}
\usepackage{algpseudocode}
\usepackage{algorithm}
\usepackage{soul}

\def\BibTeX{{\rm B\kern-.05em{\sc i\kern-.025em b}\kern-.08em
    T\kern-.1667em\lower.7ex\hbox{E}\kern-.125emX}}
\begin{document}

\title{Generating Clarifying Questions for Query Refinement in Source Code Search}

\author{\IEEEauthorblockN{Zachary Eberhart and Collin McMillan}
\IEEEauthorblockA{\textit{Department of Computer Science} \\
\textit{University of Notre Dame}\\
Notre Dame, IN, USA \\
\{zeberhar, cmc\}@nd.edu}
}

\maketitle
\begin{abstract}
In source code search, a common information-seeking strategy involves providing a short initial query with a broad meaning, and then iteratively refining the query using terms gleaned from the results of subsequent searches. This strategy requires programmers to spend time reading search results that are irrelevant to their development needs. In contrast, when programmers seek information from other humans, they typically refine queries by asking and answering clarifying questions. Clarifying questions have been shown to benefit general-purpose search engines, but have not been examined in the context of code search. We present a method for generating natural-sounding clarifying questions using information extracted from function names and comments. Our method outperformed a keyword-based method for single-turn refinement in synthetic studies, and was associated with shorter search duration in human studies.

\end{abstract}


\vspace{-.2cm}
\section{Introduction}
\textbf{Source code search} (SCS) engines are systems that return lists of source code fragments that are relevant to user-provided search queries\cite{reiss2009semantics}. Programmers use SCS engines during software maintenance for feature location, code reuse, bug triage, and more~\cite{yan2020code}. However, programmers often fail to find the information they need with a single search~\cite{sadowski2015developers}. Search issues can be caused by a mismatch between the concepts and terms humans use to describe programming tasks, and those that search engines associate with relevant code~\cite{biggerstaff1994program, furnas1987vocabulary, mcmillan2011portfolio}.


When an initial search is unsuccessful, programmers can perform \textbf{query refinement}, which refers to the process of modifying a search query in order to retrieve more-relevant results. Over the course of a search session, programmers use information gleaned from previous search results to refine subsequent queries~\cite{sadowski2015developers}. Many source code search engines offer features to improve the query refinement process, such as suggesting relevant keywords~\cite{martie2017understanding}.


By contrast, when programmers seek information from other humans, they typically refine queries by asking and answering \textbf{clarifying questions}~\cite{tavakoli2021analyzing, kato2013clarifications, wood2018detecting}. Clarifying questions (CQs) are questions intended to confirm or elicit information about some aspect of a query~\cite{aliannejadi2019asking}. If a novice programmer were to ask an expert programmer for a function to ``convert a float'', the expert would be uncertain whether to suggest a function that, e.g., converts a float to an int, converts a float to a string, or converts another data type to a float. By asking a clarifying question such as ``Would you like to convert a float to an int or a string?'' or ``Would you like to convert a float, or convert something to a float?'', the expert would then be able to suggest code implementing the correct functionality.


An emerging trend in software engineering literature is the idea that tool support should emulate the kinds of support offered by human programmers~\cite{shihab2021summary, eberhart2021dialogue, zhang2020chatbot4qr,xie2018intelligent}; in particular, leading researchers have championed technology to help programmers better express their information needs in search queries~\cite{robillard2017demand}. Meanwhile, a growing body of work supports the use of CQs for query refinement across a broad range of domains~\cite{aliannejadi2019asking}, including software engineering~\cite{zhang2020chatbot4qr}. 




Nevertheless, clarifying questions remain understudied in the context of source code search. The more-general problem of query refinement in SCS has been studied extensively by Hill~\emph{et al.}~\cite{hill2009automatically}, Wang ~\emph{et al.}~\cite{wang2014active}, Treude~\emph{et al.}~\cite{treude2014extracting}, and Martie~\emph{et al.}~\cite{martie2017understanding}. Most existing approaches to query refinement rely on GUI elements (such as drop down menus or lists of tags), which are able to present numerous options for refinement simultaneously. By contrast, a CQ is communicated via natural language and must target a narrow aspect of the user's query to clarify~\cite{stoyanchev2014towards}. The challenge is that there is no clear way to identify potentially-relevant aspects of a SCS query or select which aspect to ask about.


In this paper, we propose an approach to interactively refine SCS queries using clarifying questions. Our approach discerns relevant aspects of a SCS query from the similarities and differences between the search results that query produces. We divide our approach into three components: 1) Identifying potentially ambiguous aspects of a code search query, 2) Selecting a query aspect to inquire about and generating a grammatically-appropriate CQ, and 3) Reranking the search results based on the user's answer.

Our evaluation methodology is twofold: First, we perform a synthetic evaluation using the CodeSearchNet~\cite{husain2019codesearchnet} dataset, which contains relevance ratings for code search results. This evaluation demonstrates that our approach quickly improves the rankings of relevant results. Second, we perform intrinsic and extrinsic human studies. We hire 10 programmers to rate the quality of CQs generated by our approach, and we hire 12 programmers to complete code search tasks aided by a CQ query refinement engine. We find that CQs reduce search duration compared to a keyword recommendation baseline.


\section{Background and Related Work}

\subsection{Clarifying Questions for Query Refinement}

Clarifying questions can help information-retrieval (IR) systems resolve ambiguous queries in one of two ways: by \textbf{confirming} the system's interpretation of the user's information need (e.g., ``is this what you are searching for?''), or by \textbf{eliciting} a missing piece of information (e.g., ``which of these are you searching for?'')~\cite{radlinski2017theoretical, zou2020empirical, zamani2020generating, sekulic2021towards,aliannejadi2021building,krasakis2020analysing}. 

A growing body of evidence indicates that these questions can improve user experience in IR systems; for instance, Bing users reported higher levels of confidence in their search results after answering CQs to refine their queries~\cite{zamani2020generating}. Users also have high rates of engagement with CQs~\cite{zou2020empirical}, particularly compared to query suggestions that are not formatted as questions~\cite{zamani2020generating}.
Other studies have observed CQs information-seeking conversations among programmers~\cite{wood2018detecting, gottipati2011finding, knauss2012detecting}. Gao~\emph{et al.}~\cite{gao2020technical} analyzed over 2M posts on technical Q/A sites; they found that a large number of comments on posts contain CQs, and that posts with CQs were more likely to receive a correct answer than those without. 


Despite these observations, few query refinement approaches in software engineering literature actually attempt to generate CQs. A key obstacle is that methods to generate CQs in broader domains rely on data that are not readily available for SCS. They typically work by identifying query \textbf{aspects} and query \textbf{facets} to enquire about~\cite{kong2013extracting}. A query aspect is a word or phrase describing a distinct information need relevant to a query; for example, given the query ``JPEG image'', one aspect could be \emph{ways to process JPEG}. A query facet is a set of terms sharing a semantic relationship to a query aspect; the facet corresponding to the \emph{ways to process JPEG} aspect could include \{\textit{convert}, \textit{rotate}, \textit{resize}\}. A more specific aspect would be \emph{image types to which JPEG can be converted}, with the facet \{\textit{PNG}, \textit{GIF}, \textit{TIFF}\}. Those facet terms might also apply to the aspect \emph{image types that can be converted to JPEG}. 

Aspects and facets serve as logical categories and options for query refinement; furthermore, because they are defined by semantic roles, they can be used to generate grammatically-correct CQs~\cite{kong2013extracting}. For instance, the aspect \emph{image types to which JPEG can be converted} could produce a question confirming the user's information need: ``Do you want to convert a JPEG to a different image type?''. Or, it could elicit a missing facet value for that aspect: ``To which image type would you like to convert a JPEG? a) PNG, b) GIF, c)TIFF.''

The main challenge for SCS is that these query aspects and facets are not known in advance~\cite{gu2004component}. Methods exist to extract query aspects and facets dynamically for general-purpose web search queries using a) reformulation data derived from query stream mining~\cite{zamani2020generating}, or b) semantic patterns found in the search results themselves~\cite{kong2013extracting}. The former approach requires a volume of reformulation data that is not available for SCS. The latter approach is more feasible, but must account for the fact that there may be limited quantities of natural language text associated with code snippets. A method for generating clarifying questions for SCS needs to identify relevant query aspects and facets, select ones that will allow for meaningful clarification, create a natural language question, and use the answer to improve the search results. To inform our approach, we consider other query refinement methods for SCS.

\subsection{Query Refinement in Software Engineering}

\begin{table}[]
\renewcommand{\arraystretch}{1.1}
\caption{Selection of related work in query refinement for search tasks in software engineering.  R=Relevance Feedback-based, F=Facet-based, C=Concept-based, T=Task-based. The NL column indicates methods with a natural language interface.}
\label{tab:relatedqr}
\centering
\vspace{-.1cm}
\begin{tabular}{|c|c|ccc|}
\hline
Project & Year & Domain & Method & NL\\
\hline
Gu~\emph{et al}~\cite{gu2004component} & 2004 & API & F & X\\
Shepherd~\emph{et al}~\cite{shepherd2007using} & 2007 & Code & T & \\
Gay~\emph{et al.}~\cite{gay2009use} & 2009 & Code & R & \\
Eisengerb~\emph{et al}~\cite{eisenberg2010apatite} & 2010 & API & F & \\
Hill~\emph{et al}~\cite{hill2011improving} & 2011 & Code & T & \\
Roldan-Vega~\emph{et al}~\cite{roldan2013conquer} & 2013 & Code & C, T & \\
Wang~\emph{et al}~\cite{wang2014active} & 2014 & Code & R & \\
Treude~\emph{et al}~\cite{treude2014extracting} & 2014 & API & C, T& \\
Martie~\emph{et al}~\cite{martie2017understanding} & 2017 & Code & F, C & \\
Li~\emph{et al}~\cite{li2018api} & 2018 & API & C & \\
Sivaraman~\emph{et al}~\cite{sivaraman2019active} & 2019 & Code & R & \\
Zhang~\emph{et al}~\cite{zhang2020chatbot4qr} & 2020 & Web & F & X\\
Xie~\emph{et al}~\cite{xie2020api} & 2020 & API & T & \\
Eberhart and McMillan~\cite{eberhart2021dialogue} & 2021 & API & C & X\\
\hline
\end{tabular}
\vspace{-.4cm}
\end{table}

\Cref{tab:relatedqr} summarizes key related work in query refinement for information retrieval in software engineering. We focus on four categories: \textit{relevance feedback}-based, \textit{facet}-based, \textit{concept}-based, and \textit{task}-based methods.

Relevance feedback-based methods simply ask users to rate individual search results as relevant or irrelevant. Gay~\emph{et al.}~\cite{gay2009use} and Wang~\emph{et al.}~\cite{wang2014active} implement relevance feedback using the Rocchio algorithm~\cite{rocchio1971relevance} to update a vector representation of the search query and rerank the results. 

Facet-based methods let users filter the search results by selecting facet values for predefined query aspects. In SCS, these aspects are limited to explicitly-defined properties of the source code (e.g., return type, parameter types, parent class). Zhang~\emph{et al.}~\cite{zhang2020chatbot4qr} proposed a facet-based method to generate targeted clarifying questions for StackOverflow post retrieval. However, this method relied on existing technical tags that SO users had assigned to the posts, and required the authors to manually identify 20 query aspects and categorize over 3700 tags into corresponding facets.


\begin{figure*}[]
	\centering
	\includegraphics[width=\textwidth]{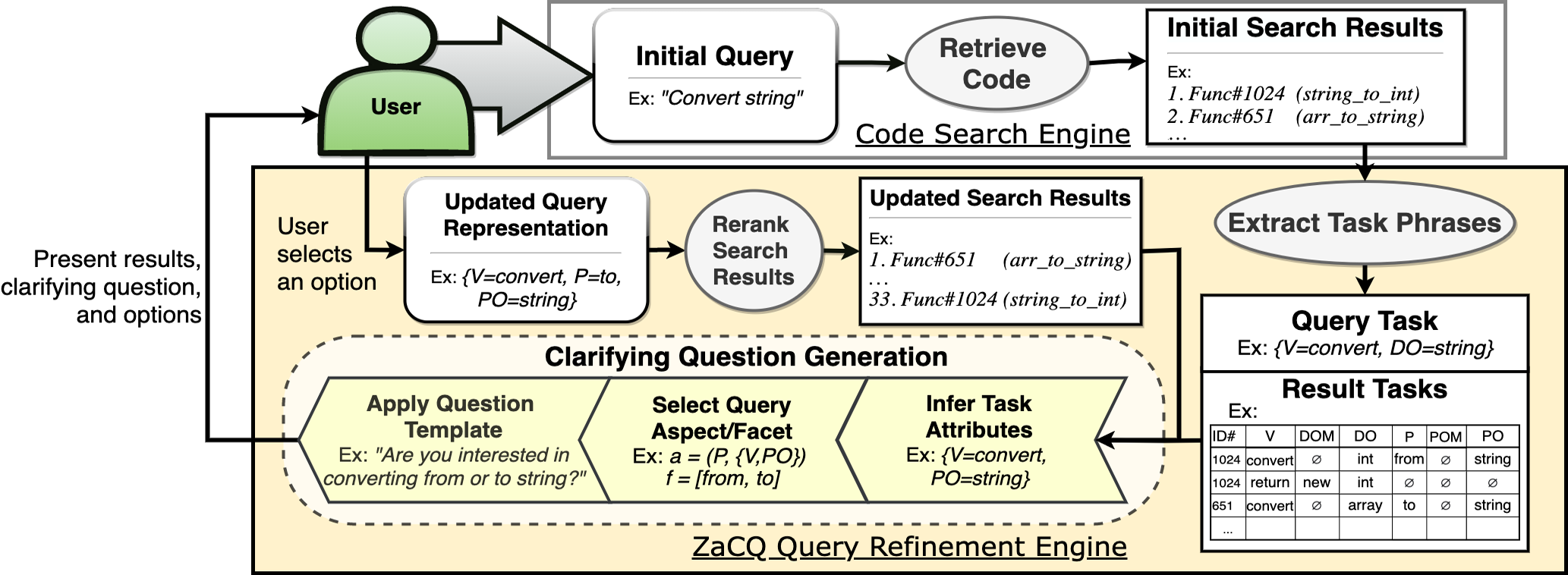}
\vspace{-.55cm}
	\caption{Overview of the ZaCQ method for natural language query refinement.}
    \label{fig:overview}
    \vspace{-.55cm}
\end{figure*}

Concept-based methods overcome these limitations by extracting discriminative features from the search results. The simplest concept-based methods are keyword recommendation algorithms. Poshyvanyk and Marcus~\cite{poshyvanyk2007combining} use Latent Semantic Indexing (LSI) to find important keywords for a set of code snippets; they then use formal concept analysis (FCA) to let users iteratively filter their searches with increasingly-specific keywords. Other methods extract certain syntactic patterns from source code or documentation to recommend entire noun phrases (e.g., ``JPEG image'')~\cite{martie2017understanding, treude2014extracting, roldan2013conquer}.

Task-based methods extend this idea to verb phrases (e.g., ``convert JPEG image to PNG''), enabling users to specify more-complex types of functionality. We refer to these functionality verb phrases as development tasks, or simply \textbf{tasks}~\cite{treude2014extracting}.  Shepherd~\emph{et al.}~\cite{shepherd2007using} first developed an approach to use verbs and direct objects extracted from function identifiers and comments for interactive query reformulation. Hill~\emph{et al.}~\cite{hill2011improving} presented a subsequent technique to generate tasks comprising verb, noun, and prepositional phrases from code snippets, and built a code navigation tool based on the resultant semantic hierarchies. Treude~\emph{et al.}~\cite{treude2014extracting} developed a technique to extract the same task phrases from API documentation by searching for specific syntactic patterns. 

We observe that task phrases convey the similarities and differences among code snippets by way of the same structured semantic relationships that connect aspects and facets. Our approach takes advantage of this property to dynamically extract aspects and facets for clarifying question generation.

\vspace{-.1cm}
\section{Approach}
\vspace{-.1cm}

This section presents our approach: the \emph{Zero-aspect Clarifying Question} (\textbf{ZaCQ}) system. ZaCQ is a natural language query refinement engine that works alongside a standard SCS engine. It targets the scenario where a SCS engine retrieves one or more results that satisfy a user's information need, but ranks them below other, less relevant results. Given a SCS query and results, ZaCQ generates a targeted clarifying question and provides options for refinement. Importantly, \textit{ZaCQ does not rely on predefined query aspects}; instead, it derives potentially-relevant aspects and corresponding facets from tasks associated with the search results. Over multiple rounds of refinement, ZaCQ identifies a single development task relevant to the user's information need.







\vspace{-.1cm}
\subsection{Overview}

Figure 1 presents a high-level overview of our approach, initiated by the user in the upper-left corner. The white box to the right of the user represents a standard SCS engine. The user provides a text query to the SCS engine, which produces a ranked list of the top-$k$ most-relevant functions from a source code repository. These results serve as the input to ZaCQ. 

First, ZaCQ uses natural language processing techniques to extract tasks from the query and search results (\Cref{sec:featex}).

Second, ZaCQ generates a clarifying question based on the extracted tasks (\Cref{sec:cqgen}). In order to create a question targeting an unclear aspect of the user's information need, it attempts to infer probable task attributes (e.g., common actions or objects) from the query and results. It then selects a salient query aspect to enquire about -- either by seeking confirmation that it is relevant to the user, or eliciting a value from the corresponding facet. Based on the chosen aspect/facet, ZaCQ applies an appropriate template to create a CQ.

Third, ZaCQ presents the search results, CQ, and refinement options. The user may select an option to refine his/her query. ZaCQ creates an updated query representation based on the user's selection, which it uses to rerank the search results (\Cref{sec:rerank}). If aspects of the user's information need are still unclear, the process repeats, and the system generates a new CQ targeting a different query aspect.

\vspace{-.1cm}
\subsection{Task Extraction}
\label{sec:featex}


ZaCQ begins by generating tasks associated with a user query $q$ and the corresponding list of search results $R$. A task $t$ is represented as a data structure with string values for up to six syntactical roles: a verb (V), a direct object modifier (DOM), a direct object (DO), a preposition (P), a prepositional object modifier (POM), and a prepositional object (PO).  A value for a particular syntactic role $s$ is referred to as a task attribute $t_{s}$. Some attributes may be missing, but each task must have an value for V and either a DO value or P and PO values.

We extract tasks from $q$ and $R$ using the method published by Treude~\emph{et al.}~\cite{treude2014extracting}. This method was found to be highly accurate at extracting meaningful development tasks from software documentation. We refer readers to the original paper for algorithmic details, and to our repository (see \Cref{sec:conclusion}) for implementation details; here, we briefly summarize the method and adaptations made for our approach.

\subsubsection{Preprocessing} For each function in $R$, we attempt to extract tasks from the function name and comment (if there is one). We split camelcase and snakecase function names into separate tokens and prepend them to the comment string with a period. We then preprocess the comment, adapting the steps used by Treude~\emph{et al.} for docstring-specific text formatting.


\subsubsection{Dependency Parsing} 
We use \textit{spaCy}~\cite{spacy} to parse syntactic dependencies. Extracting a task involves following specific dependencies and recording tokens in certain syntactic roles. All tasks start with a verb; we follow dependencies to find values for DO, P, and PO roles, and then continue parsing from objects to find DOM/POM modifiers.  One modification we make is to follow prepositional dependencies on objects denoting collections of elements (e.g. ``list of ints''); we append these phrases to the corresponding $t_{DO}$ or $t_{PO}$ strings. 

\subsubsection{Postprocessing} 
Like Treude~\emph{et al.}, we filter out tasks with attributes that are too generic to make for valuable clarification targets.  We filter out a list of generic nouns, which including generic programming terms (e.g., ``parameter'', ``function''), as well as a small list of generic verbs (e.g., ``take'', ``be'', ``do'', ``have''). 


Using this approach, we attempt to extract a task from $q$, and as many tasks as possible from each function in $R$. We store tasks extracted from $R$ in a task table $T$, where each row corresponds to a single task for a single result. 

\subsection{Clarifying Question Generation}
\label{sec:cqgen}

Next, we use the task table $T$ and user query $q$ to generate a natural language CQ targeting a particular query aspect $a$. We define an aspect as a target syntactic role $s$ given a set of defined task attributes $d$; e.g., the aspect \textit{image types to which JPEG can be converted} would be represented as ($s$=POM, $d$=\{V=\textit{convert}, DO=\textit{jpeg}, P=\textit{to}, PO=\textit{image type}\}). Because query aspects are not known in advance, we use the tasks in $T$ to identify potential aspects. The set of all query aspects $A_q$ is defined as $A_q= \{(s,d)\mid s\in S_d, d\in D\}$, where $D$ is the set of all subsets of task attributes for each task in $T$ and $S_d$ is the set of all syntactic targets for $d$.

Possible syntactic targets include the six syntactic roles that define tasks, as well as three special types: 1) object (O), which targets a DO or PO, 2) object modifier (OM), which targets a DOM or POM, and 3) object role (OR), which targets a minimal verb phrase including a V and either a DO or a P and PO. The O and OM types enable ZaCQ to enquire about relevant objects without concern for the specific syntactic roles they play. The OR type allows ZaCQ to distinguish whether an O attribute serves as a DO or a PO. 

We specify a set of rules to determine $S_d$ for a given $d$. The purpose of these rules is to avoid asking users confusing or difficult CQs; e.g., when $d$ is empty, a CQ should not elicit a preposition from the user. The rules are as follows:

\begin{algorithm}
\begin{algorithmic}[1]
\State \textbf{if} {$d = \emptyset$} \textbf{then} $S_d\gets{\{\text{V}, \text{O}\}}$
\State \textbf{else if} {$\text{V}\not\in d.\text{roles}$} \textbf{then} $S_d\gets{\{\text{OM}, \text{OR}\}}$ 
\State \textbf{else} $S_d\gets{\{\text{DO}, \text{P}, \text{PO}\}}$ \textbf{end if}


\State \textbf{if} $\text{DO}\in d.\text{roles}$ \textbf{then} $S_d\gets S_d\cup {\{\text{DOM}\}}$ \textbf{end if}
\State \textbf{if} $\text{PO}\in d.\text{roles}$  \textbf{then} $S_d\gets S_d\cup {\{\text{POM}\}}$ \textbf{end if}
\State $S_d \gets \{S_d \setminus d.\text{roles}\}$
\end{algorithmic}
\end{algorithm}

ZaCQ can generate two kinds of CQs for an aspect: CQs confirming the aspect's relevance to the user's information need, and CQs eliciting a value for the aspect's syntactic target $s$. When eliciting a value, ZaCQ presents up to $n$ options from the corresponding query facet $f_a$. The facet comprises the set of unique $t_s$ attributes in $T_d$, where $T_d=\{t\in T \mid d \subseteq t\}$. The type of CQ generated depends on the number of options presented: $\leq1 \rightarrow \text{confirmation}$, $\geq2 \rightarrow \text{elicitation}$.

In order to select an appropriate query aspect/facet for a CQ, ZaCQ follows a hand-crafted refinement strategy. First, ZaCQ chooses a set of defined attributes $d$ for a query aspect. A cautious refinement strategy would be to use a $d$ comprising only the attributes that the user had previously accepted. However, this may not be the most efficient strategy, as it can lead to ZaCQ explicitly clarifying attributes that should be obvious in context, e.g. an attribute that appeared in the task extracted from the user query, or one that appears in tasks in the majority of search results. ZaCQ balances caution and efficiency by inferring task attributes from $T$ and $q$.



\subsubsection{Attribute Inference}
The goal of attribute inference is to populate $d$ with attributes that are likely to be relevant to the user's information need. We use the frequency of an attribute in $T_d$ as a proxy for relevance; e.g., an attribute that appears in tasks for 6 functions is considered twice as relevant  as one that appears in 3 functions. 

To infer attributes, ZaCQ first sets $d$ equal to the set of previously-accepted attributes and gets $S_d$. It then considers attributes in the query facet for each candidate aspect. If any of the attributes also appear in the task extracted from $q$, that attribute is inferred. Otherwise, if the most-common attribute meets a minimum support and confidence threshold (specified as hyperparameters), it is inferred. If an attribute is inferred, ZaCQ adds it to $d$ and repeats the process for the new $S_d$.

\subsubsection{Aspect/Facet Selection}
Next, ZaCQ uses a greedy algorithm to select a query aspect $a$ and facet $f_a$ to ask about. For $a$, ZaCQ selects the syntactic role $s$ in $S_d$ that has the single most-frequent attribute in $T_d$. It then selects the top-$n$ most-common attributes in $f_a$ to present as refinement options.

\subsubsection{Question Templating}
Finally, we apply a natural language CQ template to the selected aspect/facet. We define 5 templates for different syntactic targets, adapted from prior work in CQ generation for web search~\cite{zamani2020generating, wang2021template}:
\vspace{-.1cm}
\begin{enumerate}
    \item[T1)] ``Are you interested in {[verb phrase]}?''
    \item[T2)] ``Are you looking for {[object phrase]}?''
    \item[T3)] ``What kind of {[object phrase]} are you interested in {[verb phrase/none]}?''
    \item[T4)] ``How do you want to {[verb phrase]}?''
    \item[T5)] ``Found {[\#]} items related to {[object/verb phrase]}. Would you like to see them first?''
\end{enumerate} 
\vspace{-.1cm}
In elicitation questions, the target syntactical role is replaced with ``any of the following''. CQs for aspects targeting V, DO, or PO use T1; O, DO, and PO use T2; OM, DOM, and POM use T3; and P uses T4. Most confirmation questions use T1 or T2; T5 is a special case for confirming attributes inferred directly from the task extracted from $q$.

\subsection{Result Reranking}
\label{sec:rerank}
Users may respond to elicitation CQs by selecting an option or ``None.'' For confirmation CQs, users may select ``Yes'' or ``No.'' ZaCQ records all selected/confirmed attributes and sets of rejected attributes, and identifies lists of candidate functions (associated with at least one task that contains all accepted attributes and no rejected sets of attribute) and rejected functions (associated with no suitable tasks and at least one task with attributes that were explicitly rejected).

We note that SCS results may not necessarily have comments or function names from which tasks can be extracted, but may still be relevant to a query; therefore, ZaCQ does not directly filter non-candidates from the results.  Instead, it promotes all functions similar to candidate functions and demotes those similar to rejected functions using the Rocchio algorithm \cite{rocchio1971relevance}. This mechanism can be used for SCS engines that embed functions and queries in the same vector space; it works by creating an updated vector representation of the query shifted towards candidate result vectors and away from rejected ones (according to predefined hyperparameters), and then reranking each result by cosine similarity.

\vspace{-.1cm}
\section{Synthetic Evaluation}
\label{sec:quanteval}

We performed a synthetic evaluation to determine whether clarifying questions generated by the ZaCQ query refinement engine are effective at improving the relevance rankings of source code search results.  The evaluation involved generating CQs for sets of code search results, automatically selecting answers using relevance data, and recording the improvement observed in the overall ordering of the results.

We compared the ZaCQ method to two baseline methods: a method that asks users to clarify only a verb and a direct object (we refer to this method as \emph{V-DO}), and a keyword recommendation method (we refer to this method as \emph{KW}). 

\vspace{-.2cm}

\subsection{Research Questions}
We ask the following research questions ($RQ$s):


\begin{description}
\item[$RQ_1$] How well does ZaCQ perform after asking $1$ or more questions, \ul{compared to the default result ordering}?
\end{description}

The purpose of $RQ_1$ is to quantify ZaCQ's reranking performance and determine whether subsequent CQs after the first are less effective. In \cite{wang2014active}, the authors found that requesting additional relevance feedback for individual code search results yielded diminishing returns, so it is valuable to measure the utility of increasingly-specific CQs.

\begin{description}
\item[$RQ_2$] How well does ZaCQ perform after asking $1$ or more questions, \ul{compared to the baseline methods}?
\end{description}

The purpose of $RQ_2$ is to compare ZaCQ to the V-DO and KW baselines. While ZaCQ is designed to select query aspects and refinement options for natural-sounding CQs, it is also important that it be as efficient as the baselines.

\vspace{-.1cm}

\subsection{Baselines}
\label{sec:baselines}
We compared the ZaCQ method to two baselines methods: a Verb-Direct Object (V-DO) method and a Keyword (KW) method. These represent existing concept-based and task-based approaches to interactive query refinement for SCS.

The V-DO baseline is based on methods used by Shepherd~\emph{et al.}~\cite{shepherd2007using} and Hill~\emph{et al.}~\cite{hill2009automatically}. It uses the same task extraction and architecture as ZaCQ, but it can only clarify values for the verb and direct object syntactic roles. Furthermore, it always elicits the two attributes in the same order (verb 	$\rightarrow$ object); it cannot infer attributes from the search results or decide to clarify the object first. This baseline is intended to highlight whether those features in ZaCQ (detailed query aspects, inference, and decision-making) meaningfully improve its performance.



The KW baseline is based on the refinement method proposed by Poshyvanyk and Marcus~\cite{poshyvanyk2007combining}. In brief, this method uses Latent Semantic Indexing to identify 25 keywords in the search results, and it uses Formal Concept Analysis to suggest discriminative keywords over several rounds of refinement. As users select keywords, they restrict the candidate functions and subsequent keyword suggestions for that query.

We implement both baselines use the same relevance feedback-based reranking algorithm as ZaCQ.\footnote{The KW baseline uses an adapted procedure to determine candidate and rejected functions. When keywords are rejected, any functions associated with those keywords are considered to be rejected. The set of candidates comprises functions associated with any selected keywords, excluding any rejects. Rejects are accounted for when suggesting subsequent keywords.}




\subsection{Dataset}
\label{sec:dataset}
\begin{table}[!t]
    
    \caption{CodeSearchNet dataset used for the syntethetic evaluation.}
    \vspace{-.2cm}
\renewcommand{\arraystretch}{1}
\centering
\begin{tabular}{|c|cccccc|}
\hline
\rowcolor[HTML]{C0C0C0} 
Language                                                                                 & Python & Java & PHP & Ruby & Javascript & Go                        \\ \hline
\cellcolor[HTML]{EFEFEF}\begin{tabular}[c]{@{}c@{}}\# Functions\end{tabular} & 1.2M   & 1.6M & 1.0M & .2M  & 1.9M       & .7M                       \\\hline

\renewcommand{\arraystretch}{.6}
\cellcolor[HTML]{EFEFEF}\begin{tabular}[c]{@{}c@{}}\# Queries\\ in Evaluation\end{tabular}   & 55     & 33   & 8   & 7    & 6          & \cellcolor[HTML]{C0C0C0}0 \\
\hline
\renewcommand{\arraystretch}{.6}
\cellcolor[HTML]{EFEFEF}\begin{tabular}[c]{@{}c@{}} \% Results \\ with Task(s)\end{tabular}   & 57.9     &  63.3  & 66.2 & 48.8 &  34.0 & \cellcolor[HTML]{C0C0C0}N/A \\
\hline
\end{tabular}
    \vspace{-.4cm}
    \label{tab:dataset}
\end{table}

Our evaluation dataset consists of 1) a set of 99 programmer queries, 2) SCS results for those queries, and 3) relevance ratings for those search results, all of which were derived from CodeSearchNet~\cite{husain2019codesearchnet}. CodeSearchNet is a collection of datasets and benchmarks for code search evaluation. The datasets comprise about 6 million functions scraped from GitHub repositories, 2.1 million of which are paired with a comment. There are separate datasets for six programming languages: Python, Javascript, Ruby, Go, Java, and PHP. 

\subsubsection{Queries}
CodeSearchNet provides a set of 99 general natural-language programming queries, such as ``convert int to string'' and ``k means clustering''. The queries were collected from common Bing searches with high click-through rates to code, and the CodeSearchNet authors manually filtered out queries that clearly included specific technical keywords. 

\subsubsection{Search Results}
We generated the top 50 search results for each query and dataset using the state-of-the-art neural bag-of-words model packaged with CodeSearchNet. This was the best-performing model in the original CodeSearchNet publication, and remains one of the best-performing models on the CodeSearchNet benchmarks\footnote{https://wandb.ai/github/codesearchnet/benchmark/leaderboard}. We chose to generate 50 results in line with the CodeSearchNet benchmarks.

\subsubsection{Relevance Ratings}
Each dataset includes relevance ratings for SCS results for each of the 99 queries. To create these ratings, the CodeSearchNet authors used a baseline code search method to retrieve the top-10 results for each query for each dataset; they then hired programmers to rate the relevance of individual results to the corresponding query on a scale from 1 (least relevant) to 4 (most relevant). Not all results received relevance ratings, and some results with ratings do not appear in our dataset because we used a different retrieval method.


\subsubsection{Filtering}
Certain queries did not have an adequate amount of relevance data for use in our synthetic evaluation. Therefore, we filtered queries that didn't meet all three of the following criteria from the dataset:
\begin{itemize}
    \item At least three search results have ratings
    \item At least one search result has a positive rating (3 or 4) and contains both a task phrase (used by ZaCQ and V-DO) and a keyword (used by KW) 
    \item Rated search results are not already in the optimal order (i.e., ratings do not decrease monotonically with ranking)
\end{itemize}

\Cref{tab:dataset} presents the final composition of the dataset.





\subsection{Methodology}
\label{sec:quantmeth}

\begin{figure}[!t]
    \centering
    \renewcommand{\arraystretch}{.9}
    \begin{tabularx}{\columnwidth}{|lX|}
        \hline
        \textbf{Query:} & convert integer to text \\
        \textbf{Language:} & Java\\
        \textbf{Method:} & ZaCQ\\
        \hline
    \end{tabularx}
    
    \vspace{.1cm}
    
    \begin{tabularx}{\columnwidth}{ll}
        \multicolumn{2}{l}{\textbf{Initial Relevance-Annotated Results:}}\\
    \end{tabularx}
    
    \vspace{.1cm}
    
    \begin{tabularx}{\columnwidth}{llX}
        \textsc{\small{Rank}} & \textsc{\small{Score}}  & \textsc{\small{Extracted Task Phrases}}\\
        3 & 1  & [``convert text to integer'']\\
        10 & 3  & [``convert int to string value'', ``display text to screen'']\\
        24 & 4 & [``convert int to string value'']\\
    \end{tabularx}
    
    \vspace{.1cm}
    
    \begin{tabularx}{\columnwidth}{ll}
        \textbf{Reciprocal Rank:} $.10$ & \textbf{Average Precision:} $.09$  \\\textbf{NDCG:}  $.77$ &\\
    \end{tabularx}

    \vspace{.1cm}
    
    \begin{tabularx}{\columnwidth}{ll}
        \hline
    \end{tabularx}
    
    
    \begin{tabularx}{\columnwidth}{lX}
        \textbf{Clarification Aspect:} & POM $|$ V, DO, P, PO\\
        \textbf{Clarifying Question:}  & What kind of value are you interested in converting int to?\\
        \textbf{Options:}  & float, datetime, string, null\\
        \textbf{Selected Response:}  & string\\
    \end{tabularx}
    
    \vspace{.1cm}
    
    \begin{tabularx}{\columnwidth}{ll}
        \hline
    \end{tabularx}
    
    \vspace{-.1cm}
    
    \begin{tabularx}{\columnwidth}{ll}
        \multicolumn{2}{l}{\textbf{Reranked Relevance-Annotated Results:}}\\
    \end{tabularx}
    
    \vspace{.1cm}
    
    \begin{tabularx}{\columnwidth}{llX}
        \textsc{\small{Rank}} & \textsc{\small{Score}}  & \textsc{\small{Extracted Task Phrases}}\\
        1 & 3 & [``convert int to string value'', ``display text to screen'']\\
        3 & 4  & [``convert int to string value'']\\
        41 & 1  & [``convert text to integer'']\\
    \end{tabularx}
    
    \vspace{.1cm}
    
    \begin{tabularx}{\columnwidth}{ll}
        \textbf{New Reciprocal Rank:} $1$ & \textbf{New Average Precision:} $.83$  \\\textbf{New NDCG:}  $.94$ &\\
    \end{tabularx}

    \caption{Synthetic query refinement example round.}
    \label{fig:eval_sample}
    \vspace{-.4cm}
\end{figure}

Our methodology for answering the RQs involved using each query refinement method (ZaCQ, V-DO, and KW) to iteratively rerank the search results of each query in our dataset. The synthetic refinement procedures were as follows:

\textbf{1)} Given a query and a list of search results, the refinement method generated a set of up to 5 refinement options. For ZaCQ and V-DO, these options were facet terms for a single query aspect (e.g., a set of verbs, or a set of direct objects that a certain verb acts upon). For KW, these were keywords.

\textbf{2)} We simulated a user's response by selecting a relevant option. An option was ``relevant'' if it and any inferred attributes were associated with a result that was relevant to the query (i.e., rated 3 or 4). If there were multiple relevant options, the one associated with the fewest results was chosen. A response could also indicate that no relevant option was presented.

\textbf{3)} The simulated response was used to update the query representation and rerank the results.

\textbf{4)} The refinement method generated a new set of options, and repeated the process until no further refinement could occur (i.e., until ZaCQ had clarified a complete task, V-DO had clarified a V-DO pair, or KW had narrowed down a set of functions with identical keywords).

\Cref{fig:eval_sample} illustrates the synthetic evaluation's query refinement process with the ZaCQ method. Initially, the most relevant search result is poorly-ranked. ZaCQ infers from the search results that the user's information need involves converting an int to some kind of value, and chooses to clarify what kind of value the user is interested in. ``String'' is automatically selected in the evaluation because it is associated with a relevant result. ZaCQ uses this information to rerank all search results. The complete task phrase ``convert int to string value'' has been clarified, so the evaluation concludes.

\subsection{Metrics}
\label{sec:metrics}
We evaluated refinement performance using three common metrics in code retrieval and refinement literature: 1) Mean Reciprocal Rank (MRR), 2) Mean Average Precision (MAP), and 3) Normalized Discounted Cumulative Gain (NDCG). 


MRR is a metric for evaluating processes that produce lists of possible responses to a query. It considers only the rank of the first relevant (i.e., rated 3 or 4) result in a list, measuring the average reciprocal rank (the multiplicative inverse of the rank of the first relevant result) across a set of queries.

MAP is similar, but it looks at the ranks of all relevant results in the list. In other words, it rewards a system for ranking multiple relevant items highly. 

NDCG considers the relevance ratings and ranks of all rated results, such that highly ranked relevant results are rewarded, while highly ranked irrelevant results are penalized (and vice versa). Typically NDCG is calculated for all results in a list of results. However, to account for the sparse relevance data in the CodeSearchNet dataset, we calculate NDCG over the subset of results with relevance ratings.



\subsection{Results}

\begin{table}[]
\renewcommand{\arraystretch}{1.1}
\caption{Results of the synthetic evaluation. The best results for each round are bolded. Results produced by ZaCQ have asterisks for each baseline method they significantly outperform.}
\centering
\begin{tabular}{|cr|ccllll|}
\hline
\rowcolor[HTML]{C0C0C0} 
\cellcolor[HTML]{C0C0C0}                         & \multicolumn{1}{c|}{\cellcolor[HTML]{C0C0C0}}                                                                              & \multicolumn{6}{c}{\cellcolor[HTML]{C0C0C0}\# Rounds Refinement}                                                                                                                                                                                                                                                                                       \\ \cline{3-8} 
\rowcolor[HTML]{C0C0C0} 
\multirow{-2}{*}{\cellcolor[HTML]{C0C0C0}Metric} & \multicolumn{1}{c|}{\multirow{-2}{*}{\cellcolor[HTML]{C0C0C0}\begin{tabular}[c]{@{}c@{}}Refinement\\ Method\end{tabular}}} & 0                                                & 1                                                & \multicolumn{1}{c}{\cellcolor[HTML]{C0C0C0}2}             & \multicolumn{1}{c}{\cellcolor[HTML]{C0C0C0}3}             & \multicolumn{1}{c}{\cellcolor[HTML]{C0C0C0}4}             & \multicolumn{1}{c}{\cellcolor[HTML]{C0C0C0}\textgreater{}=5} 
\\ \hline
\rowcolor[HTML]{EFEFEF} 
\cellcolor[HTML]{EFEFEF}                         & Keyword                                                                                                                    & .800                                             & .876                                             & \multicolumn{1}{c}{\cellcolor[HTML]{EFEFEF}\textbf{.915}} & \multicolumn{1}{c}{\cellcolor[HTML]{EFEFEF}.917}          & \multicolumn{1}{c}{\cellcolor[HTML]{EFEFEF}.917}          & \multicolumn{1}{c|}{\cellcolor[HTML]{EFEFEF}\textbf{.917}}    \\
\cellcolor[HTML]{EFEFEF}                         & V-DO                                                                                                                       & .800                                             & \textbf{.879}                                    & \multicolumn{1}{c}{.914}                                  & \multicolumn{1}{c}{\textbf{.918}}                         & \multicolumn{1}{c}{\textbf{.918}}                         & \multicolumn{1}{c|}{\textbf{.917}}                            \\
\rowcolor[HTML]{EFEFEF} 
\multirow{-3}{*}{\cellcolor[HTML]{EFEFEF}NDCG}   & ZaCQ                                                                                                                       & \multicolumn{1}{l}{\cellcolor[HTML]{EFEFEF}.800} & \textbf{.879}                                    & .892                                                      & .907                                                      & .906                                                      & .913                                                         \\ \hline
\cellcolor[HTML]{EFEFEF}                         & Keyword                                                                                                                    & \multicolumn{1}{l}{.590}                         & \multicolumn{1}{l}{.684}                         & .872                                                      & \textbf{.971}                                             & \textbf{.971}                                             & \textbf{.971}                                                \\
\rowcolor[HTML]{EFEFEF} 
\cellcolor[HTML]{EFEFEF}                         & V-DO                                                                                                                       & \multicolumn{1}{l}{\cellcolor[HTML]{EFEFEF}.590} & \multicolumn{1}{l}{\cellcolor[HTML]{EFEFEF}.701} & \textbf{.898}                                             & .938                                                      & .943                                                      & .943                                                         \\
\multirow{-3}{*}{\cellcolor[HTML]{EFEFEF}MRR}    & ZaCQ                                                                                                                       & \multicolumn{1}{l}{.590}                         & \multicolumn{1}{l}{\textbf{.765\**\**}}             & .843                                                      & .925                                                      & .943                                                      & .967                                                         \\ \hline
\rowcolor[HTML]{EFEFEF} 
\cellcolor[HTML]{EFEFEF}                         & Keyword                                                                                                                    & .437                                             & .494                                             & \multicolumn{1}{c}{\cellcolor[HTML]{EFEFEF}\textbf{.634}} & \multicolumn{1}{c}{\cellcolor[HTML]{EFEFEF}\textbf{.692}} & \multicolumn{1}{c}{\cellcolor[HTML]{EFEFEF}\textbf{.692}} & \multicolumn{1}{c|}{\cellcolor[HTML]{EFEFEF}.692}             \\
\cellcolor[HTML]{EFEFEF}                         & V-DO                                                                                                                       & .437                                             & .503                                             & \multicolumn{1}{c}{.631}                                  & \multicolumn{1}{c}{.666}                                  & \multicolumn{1}{c}{.669}                                  & \multicolumn{1}{c|}{.670}                                     \\
\rowcolor[HTML]{EFEFEF} 
\multirow{-3}{*}{\cellcolor[HTML]{EFEFEF}MAP}    & ZaCQ                                                                                                                       & \multicolumn{1}{l}{\cellcolor[HTML]{EFEFEF}.437} & \textbf{.540\**\**}                                 & .593                                                      & .659                                                      & .673                                                      & \textbf{.700\**}     \\
\hline
\end{tabular}
\label{tab:synthresults}
\vspace{-.3cm}
\end{table}

\Cref{tab:synthresults} summarizes the results of the synthetic evaluation. All three refinement methods (V-DO, KW, and ZaCQ) were evaluated using the same set of queries and search results from the CodeSearchNet dataset, and their performance was measured in terms of MRR, MAP, and NDCG. The metrics were calculated for the results of the initial query, and then recalculated after each round of refinement. We used grid search to explore different hyperparameter configurations for each method; \Cref{tab:synthresults} presents the results of the best-performing configurations after each round of refinement.

\smallskip
\subsubsection{$RQ_1$ (Reranking Effectiveness)}
After asking a single CQ, ZaCQ improved the MRR, MAP, and NDCG of the search results by $24\%$, $30\%$, and $10\%$, respectively. These improvements are statistically significant according to a two-sided Wilcoxon signed-rank test (p-value $<.05$). Asking additional CQs allowed for further improvement, though the average improvements were attenuated. The greatest total improvements ($64\%$ MRR, $60\%$ MAP, and $14\%$ NDCG) were achieved after fully clarifying all queries.



\smallskip
\subsubsection{$RQ_2$ (Comparison to Baselines)}
Like ZaCQ, the V-DO and KW baselines improved upon the initial ordering of the search results. The greatest single-turn improvements were seen in the first round of refinement, with subsequent rounds yielding attenuated improvements.  After one round of refinement, ZaCQ significantly outperformed both baselines in terms of MRR and MAP, and tied V-DO in terms of NDCG. In subsequent rounds, both baseline methods significantly outperformed ZaCQ in terms of all metrics. After about 4 rounds, ZaCQ started to catch up with the baselines, and when all three methods are allowed to run to completion, ZaCQ did not perform significantly worse than either of the baselines.

We attribute this pattern to ZaCQ's ability to infer task attributes. Consider the query ``buffered file reader read text'' and the initial refinement options presented by each method:

\begin{table}[h!]
    \centering
    \renewcommand{\arraystretch}{1.1}
    \vspace{-.2cm}
    \begin{tabular}{c|c|l}
         Method & Target & Options \\ \hline
         V-DO & V & [``read'', ``parse'', ``return'', ``set'', ``open'']\\
         KW & Keyword & [``character'', ``input'', ``line'', ``occur'', ``stream'']\\
         ZaCQ & V, DO, P, PO & ["read text from file'']\\
    \end{tabular}
    \label{tab:my_label}
    \vspace{-.2cm}
\end{table}

ZaCQ correctly inferred all four task attributes, allowing the refinement to conclude after just one round. For the other methods, additional refinement was necessary. ZaCQ inferred attributes most frequently in the first round of refinement; when successful, these inferences allowed ZaCQ to quickly narrow down the set of candidate results, accounting for improved first-round performance. In cases where ZaCQ's inferences were incorrect, the system had to spend additional rounds establishing the correct attributes. The V-DO and KW baselines lagged behind the first-round improvements of ZaCQ, but achieved their greatest improvements sooner by avoiding wasted rounds of inference.

The total improvement different methods achieved reflected the degree to which they were able to narrow the list of candidate results. ZaCQ was designed to clarify a complete task phrase, comprising 6 syntactic roles. In the sets of 50 search results used for the evaluation, it was rare for multiple results to share identical task phrases, meaning that ZaCQ was often able to identify a single relevant function. The V-DO baseline only clarified two semantic roles, allowing for more irrelevant results to wind up in the final list of candidates.  By contrast, the KW baseline was able to narrow its candidate set down to a single relevant function for most queries, allowing it to achieve the highest final MRR.


\subsection{Threats to Validity}
\label{sec:ttvquant}
As in any study, this evaluation carries a number of threats to validity; we have taken steps to acknowledge and address these threats where appropriate. The queries, search results, and relevance scores used present threats to internal and external validity. We mitigated these threats by choosing a large, standard dataset for code search tasks with fairly reliable relevance annotations (Cohen's $\kappa$ = 0.47). Other threats include the selection of evaluation parameters. We limited the number of refinement options to $5$, emulating Zamani~\emph{et al.}~\cite{zamani2020generating}, and the number of keywords to $25$, which was the maximum number investigated by Poshyvanyk and Marcus\cite{poshyvanyk2007combining}. Selecting different parameters may have affected the observed results. Reporting the results of only the best-performing hyperparameter configurations after each round of refinement may be seen as a threat to construct validity, as that information is only available post hoc. We note that it is valuable to be able to specifically target the highest expected performance after $n$ rounds, particularly for the first round. Further research is necessary to dynamically optimize hyperparameters throughout multiple rounds of refinement.

\section{Human Studies}
We performed two human studies: a study to evaluate the \emph{intrinsic} quality of the clarifying questions generated using the ZaCQ system, and a study of the \emph{extrinsic} utility of CQs to real programmers during code search tasks. In the intrinsic study, we hired 10 annotators to rate the intrinsic quality of CQs generated by ZaCQ. In the extrinsic study, we hired 12 programmers to complete code search tasks, assisted by either ZaCQ or the Keyword recommendation baseline. 

\subsection{Research Questions}
\subsubsection{Intrinsic Study}
For the intrinsic human study, we asked the following research questions:

\begin{description}
\item[$RQ_3$] Does ZaCQ generate meaningful questions?
\end{description}

\begin{description}
\item[$RQ_4$] Does ZaCQ generate natural questions?
\end{description}

\begin{description}
\item[$RQ_5$] Does ZaCQ generate grammatical questions?
\end{description}

\begin{description}
\item[$RQ_6$] Does ZaCQ generate logical questions?
\end{description}

\begin{description}
\item[$RQ_7$] Do users prefer clarifying questions generated by ZaCQ over keyword recommendations?
\end{description}

The purpose of $RQ_3$-$RQ_6$ is to evaluate the intrinsic properties that make for a ``good'' CQ. These quality metrics are based on an analysis of CQs by Stoyanchev~\emph{et al.}~\cite{stoyanchev2014towards}. ``Meaningful'' means that a CQ seeks to make a meaningful distinction; ``Natural'' means that a CQ sounds like something a human might ask; ``Grammatical'' means that a CQ is grammatically-correct; and ``Logical'' means that a CQ logically targets a missing piece of information.

The purpose of $RQ_7$ is to compare users' first impressions of CQs generated by ZaCQ to keyword recommendations.

\subsubsection{Extrinsic Study}
For the extrinsic study, we asked the following research questions:

\begin{description}
\item[$RQ_8$] Do users engage more with clarifying questions generated by ZaCQ than keyword recommendations?
\end{description}

\begin{description}
\item[$RQ_9$] Does ZaCQ help users find relevant results faster than the Keyword baseline?
\end{description}

\begin{description}
\item[$RQ_{10}$] Does ZaCQ help users feel more confident in their search results than the Keyword baseline?
\end{description}

The purpose of $RQ_8$ is to examine whether users are more likely to engage with (i.e., click on) CQ-based or keyword-based refinement options. Although the keyword-based method can theoretically present a more discriminative set of refinement options, we hypothesize that a user-friendly natural-language interface will encourage more engagement

The purpose of $RQ_9$ and $RQ_{10}$ is to investigate the utility of CQs generated by ZaCQ during code search. We examine search duration and user confidence, as CQs have been shown to improve both metrics in general-domain search~\cite{zamani2020analyzing}.


\subsection{Intrinsic Study Methodology}

Our methodology to answer $RQ_3$-$RQ_7$ involved creating a survey asking programmers to rate the CQs generated by ZaCQ. As in the synthetic evaluation (see \Cref{sec:dataset}), we used queries and search results from CodeSearchNet. We generated CQs for all 99 CodeSearchNet queries, using search results from the Java dataset. For each query, we generated a single question with ZaCQ and KW; that is, we did not examine subsequent questions after an initial round of refinement. We chose to only look at the first question because our synthetic evaluation found that the first question yielded the greatest improvement in result reranking.

We hired $10$ participants on the crowdsourcing website \textit{Prolific.co} to complete the survey. Participants were prescreened for English fluency. Because this survey did not require participants to read or understand code, it fell in line with literature tasking non-programmers with reading software documentation~\cite{eberhart2020automatically}. Nevertheless, we limited the participant pool to participants reporting experience with ``Computer Programming.'' We also excluded participants who failed attention check questions throughout the survey.

Survey participants were first shown task instructions and examples of appropriate ratings. After completing a training exercise, they were brought to the main rating interface, which consisted of four elements: first, a text box displaying a programmer query and the corresponding CQ; second, a box containing four input fields asking the user to rate the CQ on a 1-5 Likert scale in terms the four quality criteria; third, a text box introducing the refinement prompt and options generated by the KW baseline; and finally, an input field asking the participant to indicate which refinement method they preferred.  

\subsection{Intrinsic Study Results}

\begin{table*}[]
\renewcommand{\arraystretch}{1.1}
\centering
\caption{The highest- and lowest-rated clarifying questions in the intrinsic evaluation.}
\vspace{-.1cm}
\begin{tabularx}{\textwidth}{|l|X|X|}
\hline
\rowcolor[HTML]{C0C0C0} 
Metric                                                                                                     & Highest-Rated CQ                                                                                                                                                          & Lowest-Rated CQ                                                                                                                    \\ \hline
\cellcolor[HTML]{EFEFEF}                                                                                   & \textbf{Query}:``convert string to number"                                                                                                                                          & \textbf{Query}:``find int in string"                                                                                                         \\
\multirow{-2}{*}{\cellcolor[HTML]{EFEFEF}Meaningful}                                                       & \textbf{CQ}: Found 1 function that specifically mentions converting string to number. Would you like to see it first? (Avg. score = 4.6)                                           & \textbf{CQ}:Are you interested in finding in any of the following: half of the node, range, range start, right, or text? (Avg. score = 1.7) \\  \hline
\cellcolor[HTML]{EFEFEF}                                                                                   & \textbf{Query}:``convert string to number"                                                                                                                                          & \textbf{Query}:``encode url"                                                                                                                 \\
\multirow{-2}{*}{\cellcolor[HTML]{EFEFEF}Natural}                                                          & \textbf{CQ}: Found 1 function that specifically mentions converting string to number. Would you like to see it first? (Avg. score = 4.7)                                           & \textbf{CQ}: Are you looking for any of the following: characters, method, string, unsupportedencodingexception, or url? (Avg. score = 3.1) \\  \hline
\cellcolor[HTML]{EFEFEF}                                                                                   & \textbf{Query}:``unzipping large files"                                                                                                                                             & \textbf{Query}:``encode url"                                                                                                                 \\
\multirow{-2}{*}{\cellcolor[HTML]{EFEFEF}Grammatical}   & \textbf{CQ}: Are you interested in doing any of the following: copying files, extracting files, handling files, overwriting files, or reading files? (Avg. score = 4.6)            & \textbf{CQ}: Are you looking for any of the following: characters, method, string, unsupportedencodingexception, or url? (Avg. score = 3.1) \\  \hline
\cellcolor[HTML]{EFEFEF}                                                                                   & \textbf{Query}:``priority queue"                                                                                                                                                    & \textbf{Query}:``find int in string"                                                                                                         \\
\multirow{-2}{*}{\cellcolor[HTML]{EFEFEF}Logical}                                                          & \textbf{CQ}: Are you interested in doing any of the following: changing priority, getting priority, removing priority, returning priority, or setting priority? (Avg. score = 4.6) & \textbf{CQ}: Are you interested in finding in any of the following: half of the node, range, range start, right, or text? (Avg. score = 1.6)   \\
\hline
\end{tabularx}
\vspace{-.4cm}
\label{tab:highlow}
\end{table*}

\begin{figure}[!t]
	\centering
	\includegraphics[width=85mm]{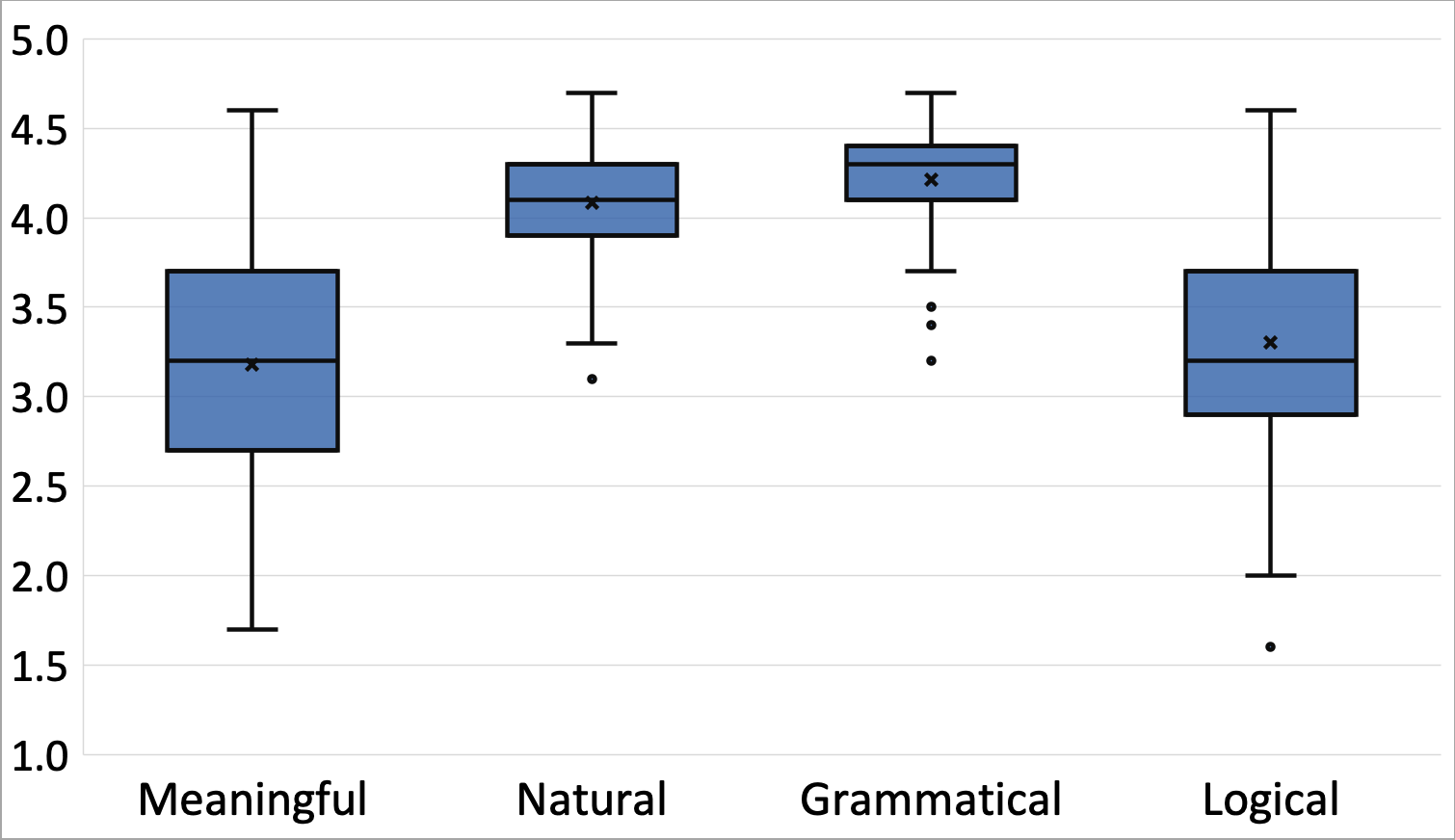}
    \vspace{-.2cm}
	\caption{Average quality ratings for ZaCQ clarifying questions in the intrinsic human study.}
    \label{fig:intrinsic_results}
    \vspace{-.3cm}
\end{figure}


We collected 10 user surveys rating 99 CQs generated by ZaCQ. We analyzed interrater reliability using intraclass correlation (ICC(3,k)). Reliability was fair to good for all quality attributes; the ``Logical'' and ``Meaningful'' qualities saw the highest levels of agreement ($r$=$.74$, $95\%$ CI ${[.65, .81]}$ and  $r$=$.76$, $95\%$ CI ${[.68, .82]}$, respectively), while the ``Natural'' and ``Grammatical'' qualities had lower agreement ($r$=$.41$, $95\%$ CI ${[.22, .57]}$ and  $r$=$.51$, $95\%$ CI ${[.35, .64]}$).


\Cref{fig:intrinsic_results} summarizes the quality ratings that participants provided, addressing $RQ_3$-$RQ_6$. For most questions, the ``Natural'' and ``Correct'' qualities were rated highly; this consistency is not surprising, given that the questions were generated using a limited set of hand-crafted templates. The ``Meaningful'' and ``Logical'' qualities saw more variation from question to question. \Cref{tab:highlow} shows the CQs that received the highest- and lowest- average ratings for each quality metric.  In general, confirmation questions (i.e., questions confirming inferred task attributes) were highly rated across all 4 metrics, while questions in which the system failed to extract task attributes from the user's query and those where semantic roles in the options appeared inconsistent received lower ratings.



To answer $RQ_7$, we looked at how frequently participants preferred CQs to KW recommendations.  Overall, $63\%$ of all ratings indicated a preference for the CQ, and a majority of participants preferred the CQ for $63\%$ of queries. Confirmation questions  were the most preferred ($70\%$), while elicitation questions targeting verbs were the least preferred ($54\%$).

\subsection{Extrinsic Study Methodology}

\begin{figure}[!b]
	\centering
	\includegraphics[width=.95\linewidth]{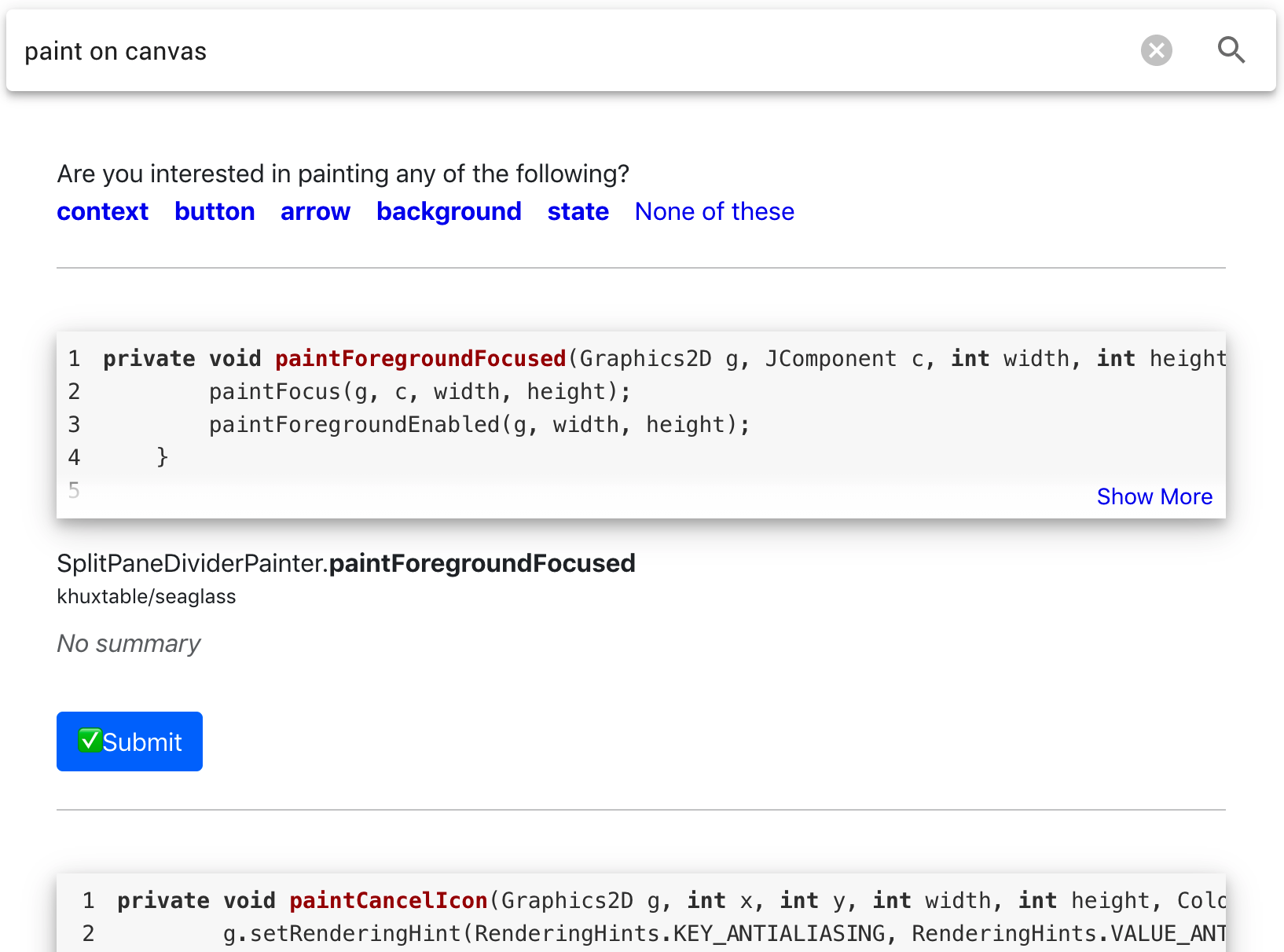}
    
	\caption{User interface for the extrinsic study. Clarifying questions and potential answers appear below the search bar.}
    \label{fig:interface}
    \vspace{-.3cm}
\end{figure}

To answer $RQ_8$-$RQ_{10}$, we ran an experiment in which programmers completed a series of code search tasks aided by either the KW or the ZaCQ refinement method. We recruited 12 programmers to participate in the experiment. All participants were Java programmers with 4 to 17 years of general programming experience (average $8.8\pm4.2$).

We designed a custom search interface (\Cref{fig:interface}) for the CodeSearchNet Java dataset (see \Cref{sec:dataset}) using neural bag-of-words search method. When users submit queries, the system retrieves the top 50 search results (displayed across 5 pages of 10 results each). The system also generates a query refinement prompt and options to display to the user. User interactions with the interface (e.g., queries, changing page, selecting refinement options) are logged in a database.

We gave programmers 8 code search tasks created by Martie~\emph{et al.}~\cite{martie2017understanding} for evaluating query refinement techniques for code search. Each task consists of a scenario, like ``You are building a sketching application'' and a request, like ``Find 4 snippets of Java source code that you think will help.'' We instructed programmers to use the interface to search as though they were actually programming a solution. When they found a function that they would want to use/replicate, we instructed them to press a ``Submit'' button. The participant would then provide confidence ratings from 1-5 indicating 1) how confident they felt that this function would \textit{correctly} address their needs, and 2) how confident they felt that they had found the \textit{best} available function for the task.

Half of the search tasks requested a single function, while the others request four. Additionally, half of the tasks requested the user find functions implementing relevant algorithms/data structure, while the rest requested any ``helpful'' functions. The combinations of response number/type pairs comprised four task categories of two tasks each. 

Like Martie~\emph{et al.}~\cite{martie2017understanding}, we used a mixed experimental design in which each participant completed four tasks using KW and four with ZaCQ. Specifically, each participant completed one task in each task category using one refinement method, and the other task in that category with the other method. We assigned refinement methods to tasks such that each task-refinement method pair was assigned to a unique set of 6 participants. As participants completed tasks, they alternated between the KW treatment and the ZaCQ treatment.  To address ordering effects, half of the participants started with ZaCQ and the other half started with KW. The order of tasks within each treatment were randomized, and participants were not explicitly informed of the different treatments.


\subsection{Extrinsic Study Results}

To answer $RQ_8$, we analyze how frequently individual participants engaged with ZaCQ and KW. To answer $RQ_8$ and $RQ_9$, we consider the search duration and confidence measures across all tasks and participants. Figures 5 and 6 summarize these results.

\subsubsection{$RQ_8$ (Engagement)}
\begin{figure*}[]
    \centering
        \begin{tabular}{cccc}
            \includegraphics[width=.22\linewidth]{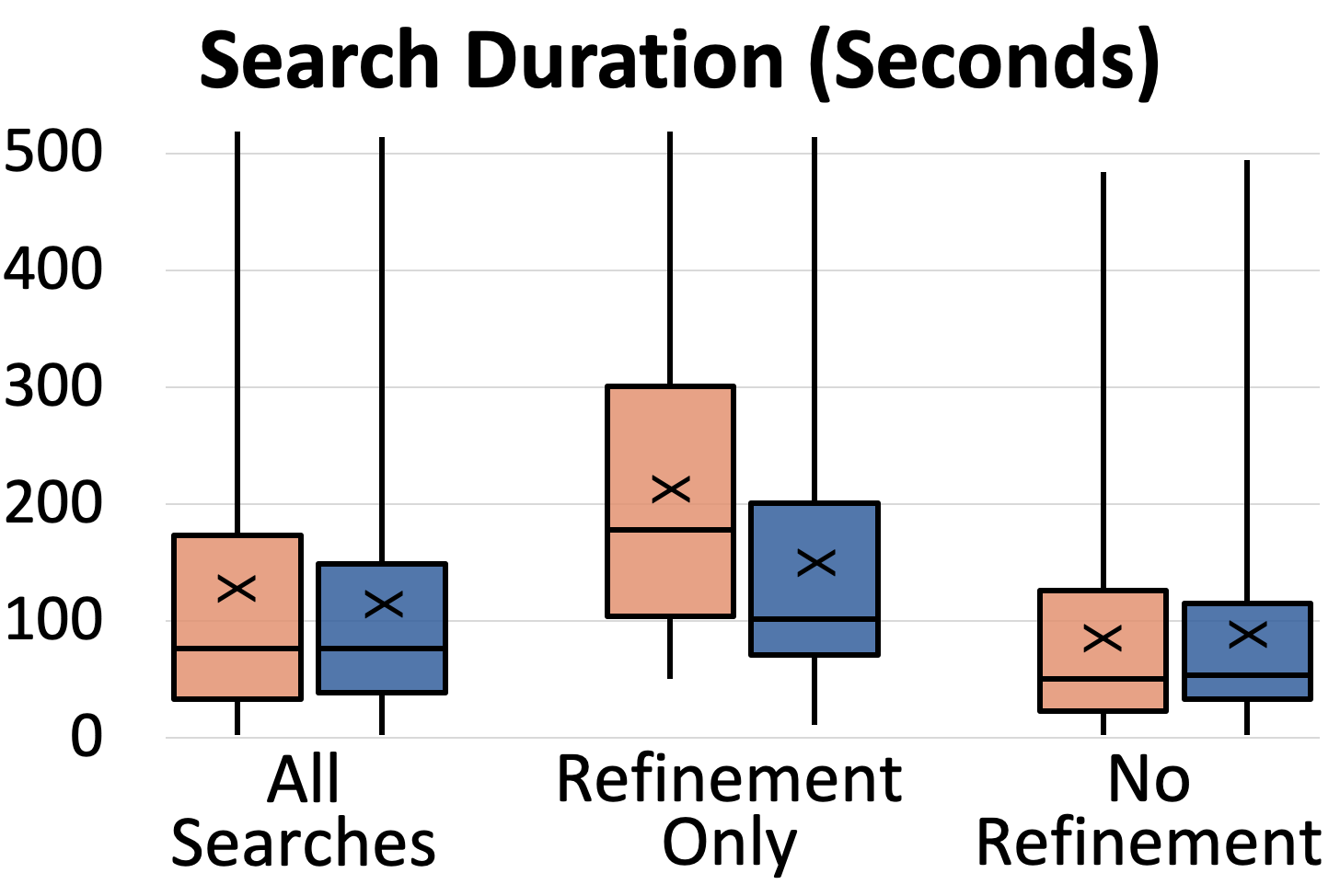} &   \includegraphics[width=.22\linewidth]{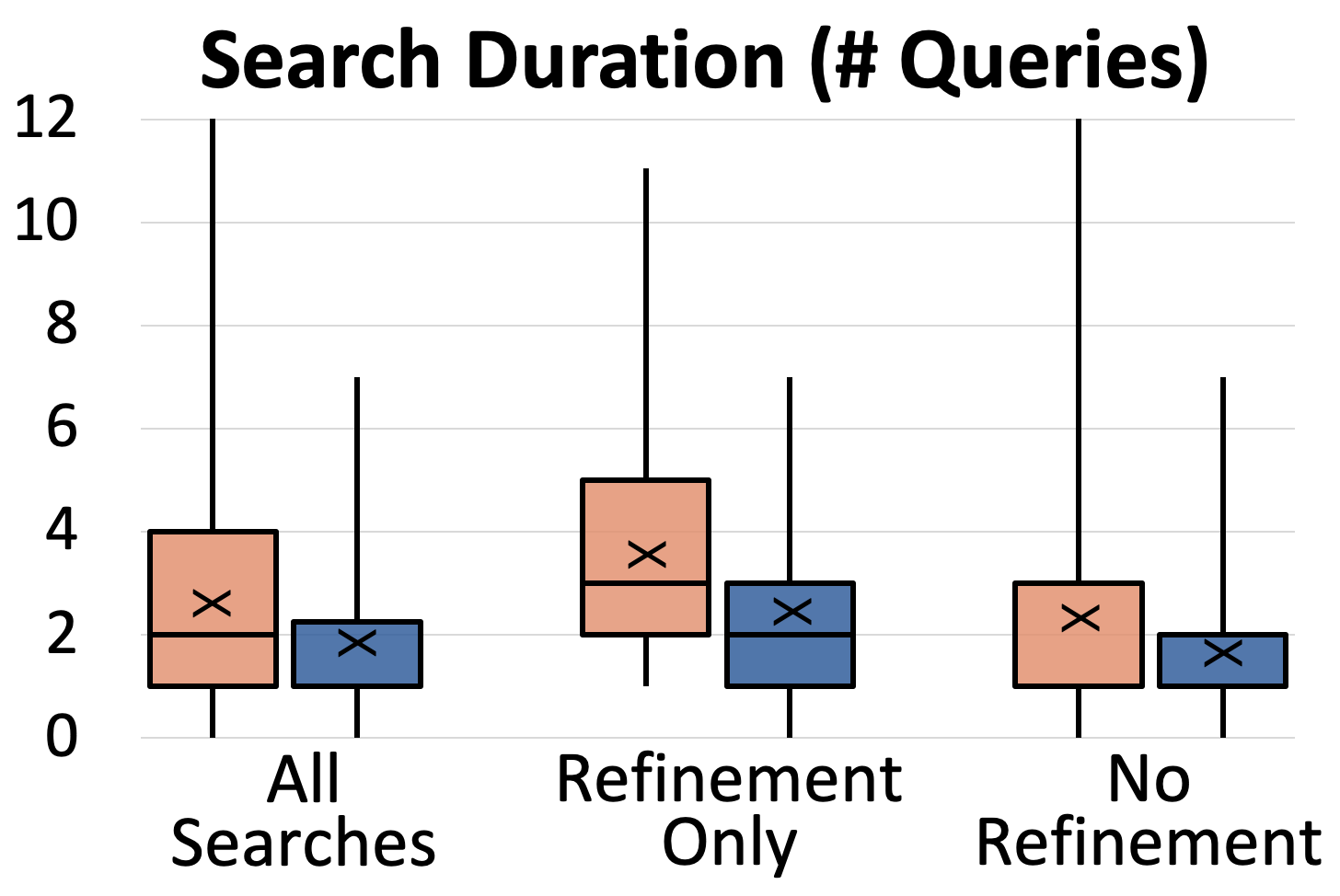} &
            \includegraphics[width=.22\linewidth]{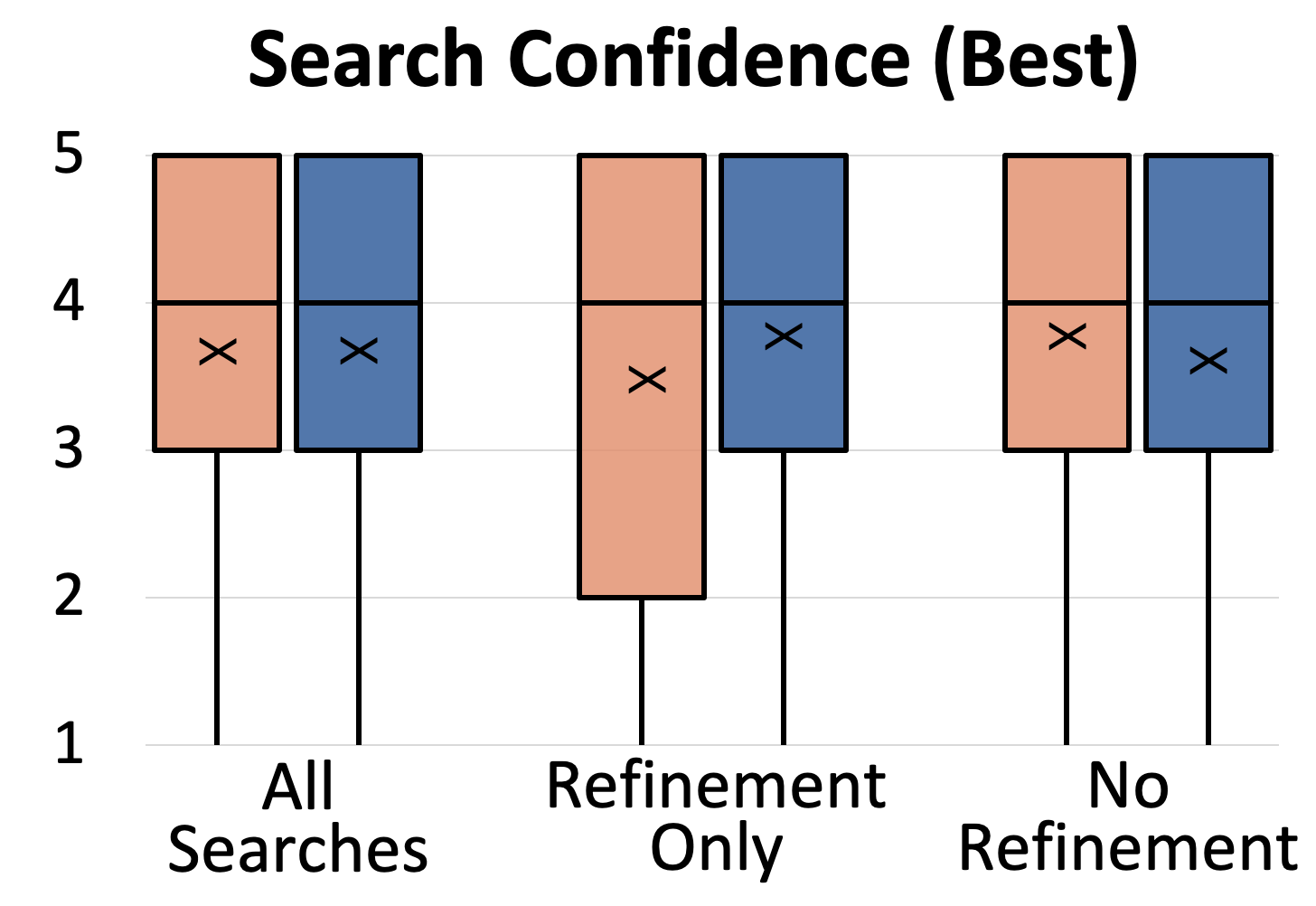} &   \includegraphics[width=.22\linewidth]{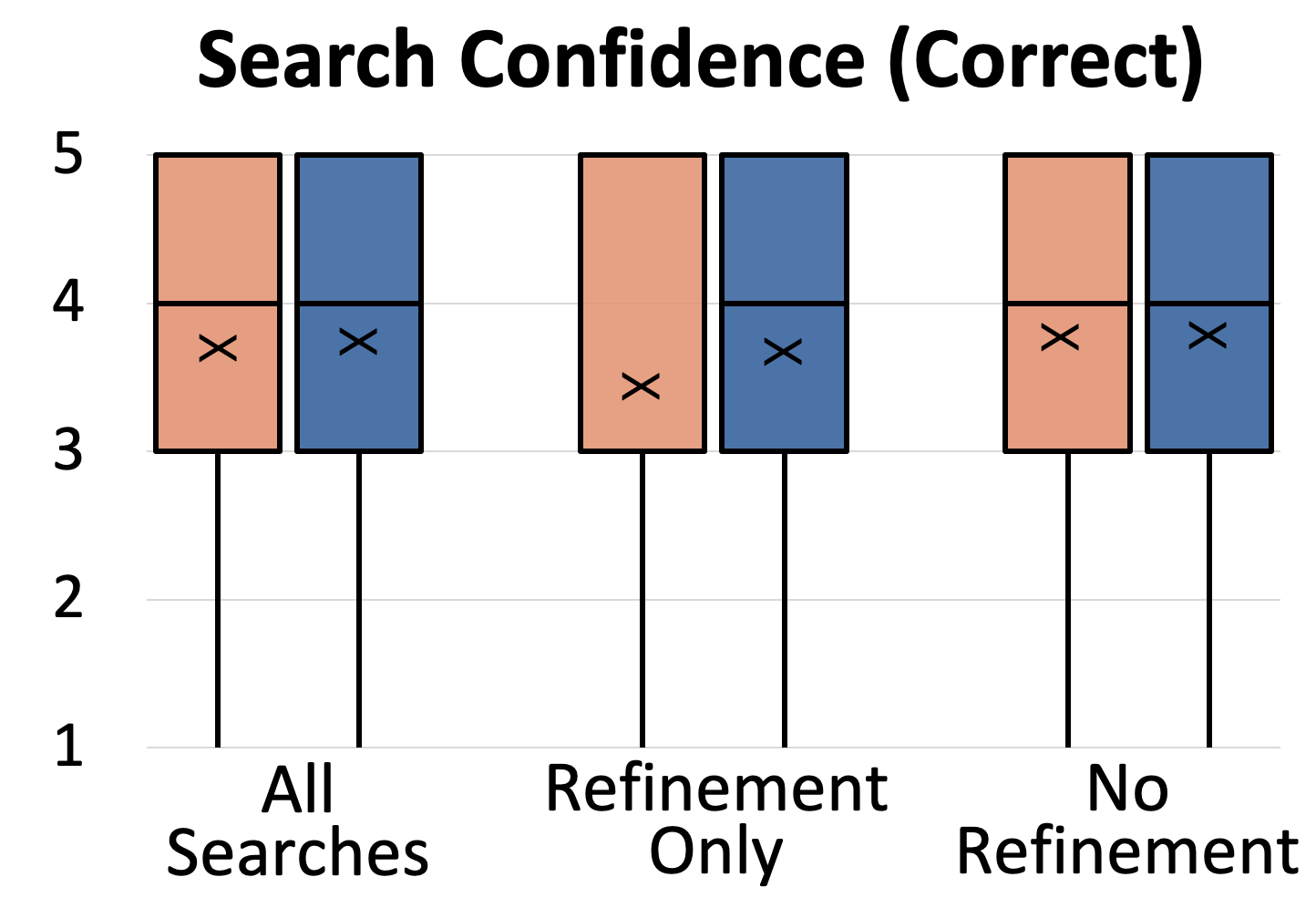} \\
        \end{tabular}
    \vspace{-0.1cm}
    \caption{\textcolor[HTML]{e09140}{$\blacksquare$}=KW, \textcolor[HTML]{2f64a0}{$\blacksquare$}=ZaCQ. Search duration and confidence ratings in the extrinsic human study. }
	\label{fig:timeconf}
	\vspace{-0.5cm}
\end{figure*}

\begin{figure}[]
	\centering
	\includegraphics[width=\linewidth]{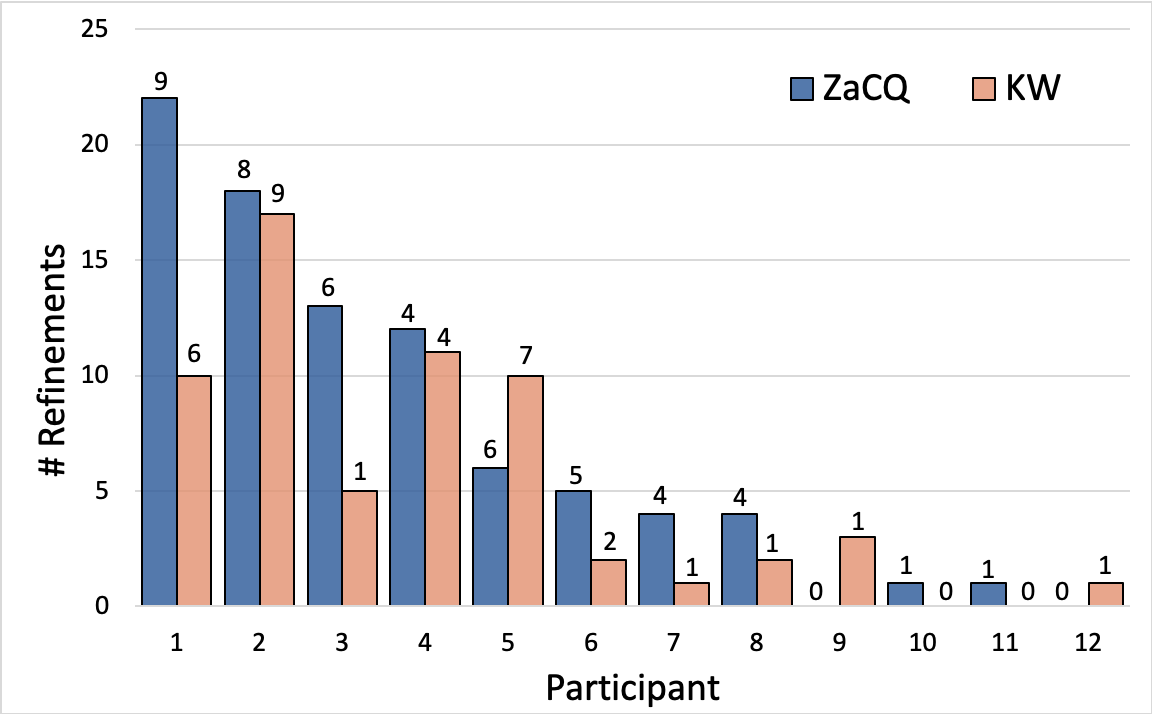}
    \vspace{-.5cm}
	\caption{Engagement in the extrinsic human study. The number above each bar indicates the number of searches in which refinement took place.}
    \label{fig:engagement}
    \vspace{-.4cm}
\end{figure}

The majority of participants engaged with more CQs than keyword recommendations. \Cref{fig:engagement} shows the total number of times each user selected a refinement option from the ZaCQ or KW interface across the 8 code search tasks. Overall, $75\%$ of the participants selected more responses to CQs than keywords. Participants selected an average of 7.2 CQ answers and 5.2 keywords over the course of the study. Furthermore, the average engagement rate per search query was $.366$ for ZaCQ and $.191$ for KW. 

We observe that 4 of the 12 participants only refined a single search query using either method. 
One of these participants reported in the exit survey that they ``often didn't even read'' the refinement suggestions throughout the study, noting that rewriting queries manually was ``just how I do things.''
Of the 8 participants who refined multiple queries, 7 of them engaged with ZaCQ more than KW.

\subsubsection{$RQ_9$ (Duration)}

We measured two metrics to evaluate $RQ_9$: the amount of time spent and the number of search queries participants used to find each submitted function (not counting time spent providing ratings/explanations). The duration of search sessions varied greatly, the quickest taking only 3 seconds and the longest 3 exceeding the 10 minute time limit. Overall, we do not observe a significant difference between the ZaCQ treatment and the KW treatment in terms of the average time and number of queries ($p>.05$ for a two-sample, two-tailed t-test). 


We note that in two-thirds of all searches, participants did not select any refinement options. To gauge the actual refinement effectiveness, we analyze the subset of search sessions in which participants selected at least one refinement option. Of the 237 total search sessions, 35 were refined by keywords and 49 were refined by CQs. We make two observations: first, the time and number of queries used for searches involving refinement were significantly higher than for searches with no refinement ($p<.01$). Second, the time and number of queries used for searches refined by CQs were significantly lower than for those refined by keywords ($p<.05$). 

These observations suggest that participants looked to the refinement interface for assistance during difficult searches. In that context, the shorter duration of searches refined by CQs suggests that CQs helped resolve difficult searches more efficiently than keyword recommendations.


\subsubsection{$RQ_{10}$ (Confidence)}

We had participants rate two measures of search confidence: confidence that an answer was correct, and confidence that an answer was the best available. As with search duration, we do not observe a significant difference in either confidence metric between the two treatments when averaging over all search sessions. The subset of searches that involved refinement have lower confidence scores than for the subset with no refinement; however, we do not observe a significant difference in search confidence between the two treatments for the refined subset ($p>.05$). 


\subsection{Threats to Validity}
Each human study carries threats to validity. For the intrinsic study, the selection of evaluation queries and search results present a threat to external validity; in particular, search results for other programming languages may have produced higher- or lower-quality questions. The selection of participants is another core threat; we aimed to reduce the threat by recruiting from a reliable crowdsourcing service~\cite{peer2017beyond} and including attention checks to filter low-quality participants.

For the extrinsic study, the selection of search tasks is a threat that may reduce generalizability. Individual differences among programmers play a role in their search behavior and preferences. We attempted to distribute the bias of individual differences by assigning a different set of programmers to complete each task with each treatment. Nevertheless, differences between programmers and tasks may bias the result averages, particularly for the subset of searches that had been refined using the experimental methods. Furthermore, as with the intrinsic study, the extrinsic study focused solely on Java source code and programmers; investigating different languages may yield different results.

\label{sec:conclusion}
\section{Conclusion}

We have presented an approach to refine source code search queries using natural language clarifying questions. Prior CQ generation methods rely on data that is not readily available for SCS. Our approach uses a task extraction algorithm to identify query aspects, and then follows a rule-based procedure for question generation. We use a feedback relevance algorithm to elevate relevant search results, including those for which descriptive task phrases are not extracted. We performed a synthetic study and two human studies to evaluate our method. It generally creates useful and natural-sounding clarifying questions; however, the inflexible rules can sometimes lead to stilted-sounding questions or repetitive options for refinement. 

Overall, we believe that CQs will play a significant role in intelligent tools for developer support. Future work should aim to incorporate more sophisticated models for result salience in the aspect inference/selection process and experiment with different result reranking algorithms.

For reproducibility, we make all source code and experimental materials available online:

\url{https://github.com/Zeberhart/ZaCQ}

\bibliographystyle{IEEEtran}
\bibliography{main}

\end{document}